\documentclass[runningheads]{llncs}
\usepackage[T1]{fontenc}
\usepackage{graphicx}
\usepackage{float}
\usepackage{times}
\usepackage{pdfpages}
\usepackage{gensymb}
\usepackage{graphicx}
\usepackage{subfig}
\usepackage{caption} 
\usepackage{xcolor}
\usepackage{amsmath}

\newcommand{\Loss}{\mathcal{L}}

\usepackage{hyperref}
\usepackage{url}
\usepackage{titlesec2}
\titlespacing*{\section}{0pt}{0.8\baselineskip}{0.4\baselineskip}
\titlespacing*{\subsection}{0pt}{0.8\baselineskip}{0.4\baselineskip}
\titlespacing*{\subsubsection}{0pt}{0.8\baselineskip}{0.4\baselineskip}

\begin{document}
\title{GaSpCT: Gaussian Splatting for Novel CT Projection View Synthesis}
\titlerunning{Abbreviated paper title}
%
\author{Emmanouil Nikolakakis\inst{1} \and
Utkarsh Gupta\inst{2} \and
Jonathan Vengosh\inst{2} \and
Justin Bui\inst{2} \and 
Razvan Marinescu \inst{2}}

\authorrunning{E. Nikolakakis et al.}
%
\institute{University of California, Santa Cruz, Electrical and Computer Engineering Department \and
University of California, Santa Cruz, Computer Science and Engineering Department
}
\maketitle              
\begin{abstract}
We present GaSpCT, a novel view synthesis and 3D scene representation method used to generate novel projection views for Computer Tomography (CT) scans. We adapt the Gaussian Splatting framework to enable novel view synthesis in CT based on limited sets of 2D image projections and without the need for Structure from Motion (SfM) methodologies. Therefore, we reduce the total scanning duration and the amount of radiation dose the patient receives during the scan. We adapted the loss function to our use-case by encouraging a stronger background and foreground distinction using two sparsity promoting regularizers: a beta loss and a total variation (TV) loss. Finally, we initialize the Gaussian locations across the 3D space using a uniform prior distribution of where the brain's positioning would be expected to be within the field of view. We evaluate the performance of our model using brain CT scans from the Parkinson's Progression Markers Initiative (PPMI) dataset and demonstrate that the rendered novel views closely match the original projection views of the simulated scan, and have better performance than other implicit 3D scene representations methodologies. Furthermore, we empirically observe reduced training time compared to neural network based image synthesis for sparse-view CT image reconstruction. Finally, the memory requirements of the Gaussian Splatting representations are reduced by 17\% compared to the equivalent voxel grid image representations.
\end{abstract}


\section{Introduction}
Implicit scene representation models such as Neural Radiance Fields (NeRF) \cite{nerf} have had a significant impact on Computer Vision and Computer Graphics applications. By mapping the coordinates $[X,Y,Z]$ of a pixel and the polar angles [$\phi$, $\theta$] describing the camera viewing direction to an RGB and transparency value, these techniques are able to implicitly capture a 3D representation into the weights of a feed-forward neural network, and can render images from novel viewing angles. One recent model for implicit 3D scene representation is Gaussian Splatting \cite{gaussiansplatting}, that uses 3D Gaussians denoted as \emph{splats} to encode view-dependent color values. These models are better suited for preserving low-detail visuals, effectively reducing both the graininess and motion-induced artifacts in the image. They require significantly less training time than other state-of-the-art NeRF models such as \cite{mipnerf360}, however their memory footprint is much larger.

Novel view synthesis and 3D scene representation would be highly desirable for use in medical imaging systems, in particular for Computational Tomography (CT). This is because the raw CT signal is acquired as projection sinograms which can then be converted into 2D angular radiographs \cite{polinadrr}. In an effort to reduce the overall dose \cite{ctrisks} administered to the patient during the scanning procedure, a typical approach is to reduce the number of projection views. Machine Learning (ML) techniques have been used for low-dose CT reconstruction. These aimed to remove statistical noise such as graininess caused by a lower amount of photons detected per projection view \cite{redcnn,ctwasserstein,cocodiff}, while other techniques dealt with sparse-view reconstruction in the 2D image domain \cite{densenetdeconvolution,framelet,ctdip} or the sinogram projection domain \cite{sinograminterp,sinogramsynth,generativesinogram}. However, since the input data consists of 2D projections, implicit representation models can potentially be used for a new form of CT reconstruction that renders novel projection views in the image domain, and which would naturally compensate for the lack of sufficient views that typically can result in structural artifacts such as streaks. MedNeRF \cite{medNerf} has adapted the GRAF \cite{graf} model which creates a conditional Generative Adversarial Network (GAN) for CT reconstruction and allows to synthesize a complete 360 degree view of a scan from a single 2D x-ray as an input. Nevertheless, there has been no work adapting Gaussian Splatting for CT novel view synthesis. 

There are several reasons why a Gaussian Splatting model is desirable for CT. First, given that radiodensities in CT imaging vary anisotropically \cite{lowdoseartifactsct}, Gaussians are an ideal way to represent a CT image. Secondly, while NeRF models reconstruct fine details and higher frequency components with higher fidelity, they can often produce floaters, which are motion-induced artifacts \cite{nerfbusters}. Gaussian Splatting is better suited for smooth images consisting of low-frequency components and is less prone to similar artifacts. Finally, training a Gaussian Splatting-based scene representation is much faster, which is critical in medical imaging, while the higher memory footprint is not a constraint in the field.

In this work, we present GaSpCT, a Gaussian Splatting-based model enhanced specifically for CT brain imaging applications. The result is a model that accurately encodes the 3D information of a CT scan given only half or less of the total projection views as a training dataset. The reconstructed 3D scenes preserve the signal with high accuracy for any camera pose within the FOV. In our experiments, the training process takes 5-10 minutes while the memory footprint of the 3-dimensional data in polygon file format (27-42 MB) is smaller than other voxel grid or mesh-based approaches. We summarize our contributions as the following:
\begin{itemize}
  \item We introduce GaSpCT, an implicit 3D scene representation and novel view synthesis model that allows for rendering novel CT brain projections from a limited projection dataset. The model leaves a small memory footprint and rendering new views is computationally inexpensive.
  \item In CT imaging, it is expected that pixels not occupied by the patient will have empty or background intensity values. To promote smoothness and sparsity in the synthesized views, we augment the baseline loss function used in Gaussian Splatting adding a Total Variation (TV) loss and Beta distribution negative log likelihood loss.
  \item We introduce a script to extract CT camera parameters from the Digital Imaging and Communications in Medicine (DICOM) metadata of the reconstructed images by approximating them to those of a pinhole camera. Thus, we remove the necessity of Structure from Motion (SfM) \cite{sfmrevisited} which performs poorly on CT radiographs due to the lack of distinct edges \cite{siftmed}. Additionally, our approach also initializes an ellipsoid 3D point cloud representing the expected patient brain volume.
  \item We perform the first verification of implicit 3D scene representation on brain CT projection images and provide all used datasets to the medical imaging community.
\end{itemize}

\section{Methodology}

\subsection{GaSpCT}
\begin{figure}[H]
    \centering
    \includegraphics[scale=0.55]{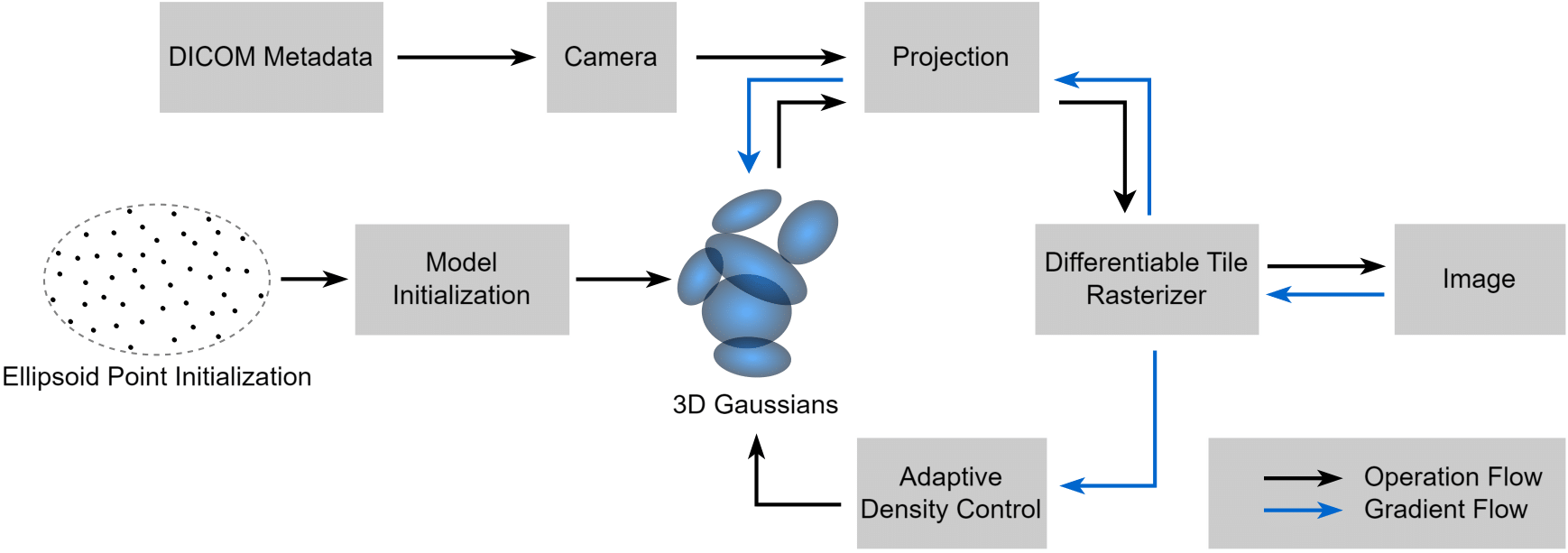}
    \caption{Optimization of 3D Gaussians initialized as an ellipsoid. Using the camera poses extracted from the DICOM metadata, we can perform forward passes and backpropagation through the differential Gaussian rasterizer.}
\label{fig:block_diagram}
\end{figure}

Our model is based on Gaussian Splatting, an implicit 3D scene representation model similar to NeRF. We adapted the model specifically for CT brain scans, adding two sparsity-promoting regularizers in the loss function and initializing the point cloud to an ellipsoid, similar to the expected brain structure seen in the training images. An overview of our model can be seen in \ref{fig:block_diagram}.

\subsection{Gaussian Splatting}

In the original Gaussian Splatting model, a 3D scene is encoded as 3D Gaussians (the splats). Each Gaussian consists of 38 parameters that encode its position, covariance, color, and opacity. During optimization, a 2D image and camera pose are sampled from the training dataset distribution. With the use of a differential Gaussian rasterizer, the equivalent image is rendered from the point cloud for the given pose. The loss between the rendered image and the ground truth is calculated and the backpropagation step is performed using the Adam optimizer on the loss function's gradients. The original Gaussian Splatting model optimizes the following loss function:

\begin{equation}
    \Loss_{Original} = (1-\lambda)\Loss_{1} + \lambda\Loss_{D\_SSIM}
\end{equation}
which is a combination of an L1 loss promoting sparsity and Dynamical Structure Similarity Index (D-SSIM) term which encourages the similarity between the rendered image and the ground truth for a given camera pose. Using this loss, the gradients of the Gaussian properties are back-propagated and the implicit representation is optimized.

\subsubsection{Total Variation Regularization}

We add a TV Regularizer in the loss function. TV penalizes large variations between neighbouring pixels and enhances smoothness in the image while also reducing the impact of noise artifacts. We implement the TV loss as done by \cite{tvrestore}:

\begin{equation}
    \Loss_{TV} = \lambda_{TV}\sum_{i,j}^{N,M}\left\vert p_{i+1,j}-p_{i,j}\right\vert + \left\vert p_{i,j+1}-p_{i,j}\right\vert
\end{equation}
where $p$ represents the pixel value at coordinates $i$, $j$. $N$ and $M$ are the image's height and width respectively. 

\subsubsection{Beta Distribution Regularizer}

We adapt the negative log-likelihood of a Beta Distribution (Beta(0.5,0.5)) as used by \cite{neuralvolumes}. This loss promotes sparsity by pushing background values to zero and enhancing the pixel intensities of the foreground.

\begin{equation}
    \Loss_{beta} =\frac{1}{P}\sum_{p}[log(I_{\alpha}(p))+ log(1-I_{\alpha}(p))]
\end{equation}
Where $P$ is the total number of pixels $p$ in the image and $I_{a}$ is the image opacity.

\subsubsection{Total Loss Function}

We combine the TV regularizer and the Beta distribution regularizer to give the total loss function:

\begin{equation}
   \Loss_{Final} = \lambda_1\Loss_{1} + \lambda_{D\_SSIM} \Loss_{D\_SSIM} + \lambda_{beta}\Loss_{beta} + \lambda_{TV}\Loss_{TV}
\end{equation}

\section{Experiments}

\subsection{Dataset}
\subsubsection{Digitally Reconstructed Radiograph}[H]
We use de-identified CT brain scans acquired from the Parkinson's Progression Markers Initiative (PPMI) study. The data, originally 3D images in DICOM format, are used as input to Plastimatch, a synthetic radiograph generation software, which generates Digitally Reconstructed Radiographs (DRR). By specifying the field of view, patient, and scanner parameters retrieved from the DICOM metadata, we simulate a CT scan using the 3D DICOM image as the input phantom. The output of the DRR is a new set of projection images. We generate 360 projection views with an angular resolution of 1 degree. The images are of 128 x 128 dimensions. In total, this procedure is used to generate DRR for 20 CT brain scans from distinct patients in order to capture anatomical variability across different subjects.

\begin{figure}[H]
    \includegraphics[width=.30\textwidth]{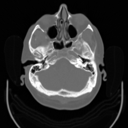}\hfill
    \includegraphics[width=.30\textwidth]{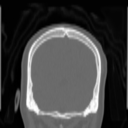}\hfill
    \includegraphics[width=.30\textwidth]{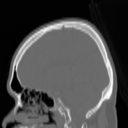}
    \\[\smallskipamount]
    \includegraphics[width=.30\textwidth]{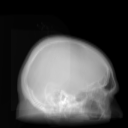}\hfill
    \includegraphics[width=.30\textwidth]{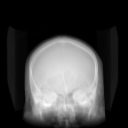}\hfill
    \includegraphics[width=.30\textwidth]{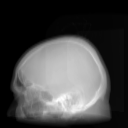}    
    \caption{Top row shows the original 3D voxel-based DICOM image retrieved from the PPMI dataset. This image is used as an input to the DRR algorithm to generate the projection image views. The bottom row shows the resulting projection views: 1) 0\degree 2) 90\degree 3) 180\degree }\label{fig:foobar}
\end{figure}

\subsubsection{Challenges with Structure from Motion on CT Images}[H]
Gaussian Splatting requires the output of SfM software  as an input to the training script. This includes the camera intrinsics and extrinsics along with the point cloud representing the identified features in the 3D scene. Nevertheless, applying SfM to CT images is particularly challenging. This is the result of radiodensity varying gradually across the reconstructed projection images. Therefore, there is an evident lack of refined edges and fine details in the image which would allow for accurate and robust feature extraction.
\subsubsection{Camera Extrinsics and Intrinsics of CT Images}[H]
Since gradient optimization for Gaussian Splatting requires the camera poses of the input images in the 3D scene, we generate the camera intrinsics and extrinsics mathematically using our prior knowledge about the CT scan parameters provided by the DICOM metadata. The variables we retrieve are the Field of View (FOV) of the imaged space, the dimensions of the detector array, and the distance of the source to the detectors and the patient. These variables are used to calculate the Cartesian coordinates $(x, y, z)$ of each camera pose. The polar angle between poses increments in the same step as the angular resolution of the CT dataset, while the azimuth angle remains fixed at 0, given we set the origin of the world coordinates in the centre of the CT FOV.

\subsection{Setup}
Our implementation is based on the open source git repository of Gaussian Splatting. All training and rendering for the model was performed on a Linux Ubuntu 20.04 Focal Fossa version. All processes were run on a single NVIDIA RTX A4000 containing a 16GB GDDR6 SDRAM. After investigating the learning loss's curve during training, we decided to run all GaSpCT tests for 20K iterations. 

\section{Results}

\begin{figure}[H]
    \includegraphics[width=.24\textwidth]{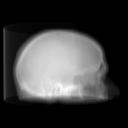}\hfill
    \includegraphics[width=.24\textwidth]{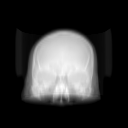}\hfill
    \includegraphics[width=.24\textwidth]{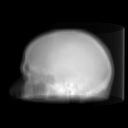}\hfill
    \includegraphics[width=.24\textwidth]{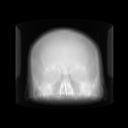}
    \\[\smallskipamount]
    \includegraphics[width=.24\textwidth]{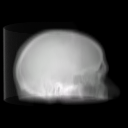}\hfill
    \includegraphics[width=.24\textwidth]{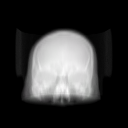}\hfill
    \includegraphics[width=.24\textwidth]{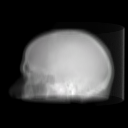}\hfill
    \includegraphics[width=.24\textwidth]{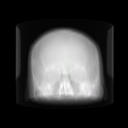}    
    \caption{Four different angular view a) 0\degree b) 90\degree c) 180\degree d) 270\degree. The top row contains the ground truth images, while the bottom row contains the rendered images for the equivalent camera poses. }\label{fig:foobar}
\end{figure}

\begin{figure}[H]
    \includegraphics[width=.24\textwidth]{Figures/rendered00000.png}\hfill
    \includegraphics[width=.24\textwidth]{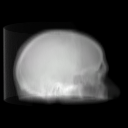}\hfill
    \includegraphics[width=.24\textwidth]{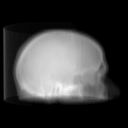}\hfill
    \includegraphics[width=.24\textwidth]{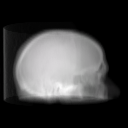}
    \caption{Four different renderings for the same projection view when testing with a) 50\% b) 25\% c) 10\% d) 5\% of the total images. }\label{fig:ratios}
\end{figure}

We ran our model on 20 image projection view scans from 20 different subjects. During training, we use 180 out of 360 images for training, and the rest for testing. We use the following metrics to evaluate the quality of the rendered images given the hidden ground truth: 1) Peak signal-to-noise Ratio (PSNR), 2) Structure Similarity Index Measure (SSIM) \cite{ssim}, 3) and Learned Perceptual Image Patch Similarity (LPIPS) \cite{lpips} using a VGG network. All metric scores can be found in \autoref{tab:baselines}. The optimization process took between 5 and 10 minutes for each of the scans. The total number of parameters optimized during the training procedure was between 4.6e+5 and 6.2e+5. Finally, the memory occupied for each output Polygon Format File was between 27-42 MB.

\begin{table}[H]
\begin{center}
\caption{Performance of GaSpCT for the brain CT dataset compared to other models using 50\% of views from the training dataset.}
\begin{tabular}{||c c c c c||}
 \hline
 Metric &  MedNeRF & MipNeRF360 & Gaussian Splatting & GaSpCT \\ 
 \hline
 PSNR $\uparrow$ & 30.36 $\pm$ 0.33 & 27.67 $\pm$ 0.23 & 40.79 $\pm$ 1.06 & \textbf{43.17 $\pm$ 1.03} \\ 
 \hline
 SSIM $\uparrow$ & 0.558 $\pm$ 0.055 & 0.14 $\pm$ 0.03 & 0.99 $\pm$ 0.002 & \textbf{0.993 $\pm$ 0.0014} \\
 \hline
 LPIPS $\downarrow$ & 0.341 $\pm$ 0.031 & 0.9 $\pm$ 0.05 & 0.017 $\pm$ 0.0037 & \textbf{0.0059 $\pm$ 0.0015} \\
 \hline
\end{tabular}
\label{tab:baselines}
\end{center}
\begin{center}
\caption{Performance of GaSpCT for various percentages of hidden images for evaluation for the brain dataset.}
\begin{tabular}{||c c c c c||}
 \hline
 Metric &  50\% of views & 25\% of views  & 10\% of views & 5\% of views  \\ 
 \hline
 PSNR $\uparrow$ & 43.17 $\pm$ 1.03  & 42.03 $\pm$ 0.95 & 38.5 $\pm$ 1.21 & 34.01 $\pm$ 1.59 \\ 
 \hline
 SSIM $\uparrow$ & 0.993 $\pm$ 0.0014 & 0.994 $\pm$ 0.0015 & 0.976 $\pm$ 0.0015 & 0.936 $\pm$ 0.017  \\
 \hline
 LPIPS $\downarrow$ & 0.0059 $\pm$ 0.0015 & 0.01 $\pm$ 0.0028 & 0.037 $\pm$ 0.0089 & 0.08 $\pm$ 0.016 \\
 \hline
\end{tabular}
\label{tab:reducedviews}
\end{center}
\end{table}

In Table \ref{tab:baselines}, we compare our metric scores to the baseline Gaussian Splatting model. Since Gaussian Splatting requires the camera poses to be provided by SfM software which fails on CT datasets due to lack of extracted features, we enhance the baseline Gaussian Splatting by automatically giving it the vector of camera poses for all CT projections. Our model outperforms the baseline Gaussian Splatting. The quantitative evaluation metrics indicate that optimizing the hyper-parameters, initializing the splats within an ellipsoid, and tailoring the loss function for CT data, collectively contribute to synthesizing 3D views that exhibit closer adherence to the ground truth observations. We trained both models for 20K iterations.

Furthermore, we compare our results to MedNeRF and MipNeRF360. To the best of our knowledge, MedNeRF is the only other open-source model that synthesizes 2D CT projection images using implicit scene representations, while MipNeRF360 allows us to compare with one of the most powerful implicit 3D scene representation methodologies. MedNeRF has a different approach to the synthesis of new projection images, as they train a conditional GAN model based on GRAF with a large collection of CT scans from different experiments to create a 3D aware latent distribution. The model then synthesizes the rest of the projections when given a single X-ray radiograph from an unseen CT scan following further iterative training of a new generator. We trained the baseline GRAF generator for 10k iterations for all our projection images from the 20 datasets. We then provided a single image from each of the 20 datasets and generated the rest of the 359 views after a further 10k iterations of training on the latent distribution. Our metric scores are superior to MedNeRF's, which is expected since GaSpCT is trained for a single 3D CT scan rather than a generalized latent distribution based on multiple scans. After training MipNeRf360, we found that the CT brain dataset failed to optimize properly following 80K iterations with a batch size of 4096 rays. It is possible that the images would be accurate if we trained the model for a longer period, but the compatibility of MipNeRF360 with CT datasets should be inspected in further detail.
Finally, we also perform a study of GaSpCT's performance when hiding a larger proportion of the data from the model during training as seen in \autoref{tab:reducedviews}. It can be seen in Fig. \ref{fig:ratios} that even when providing 5\% of the projection images, the rendered views are still very close to the ground truth. These results set the new state-of-the-art for 3D implicit scene representation for CT images and novel view synthesis for brain CT radiographs.

\section{Conclusion \& Future Work}
We have presented GaSpCT, the first implicit 3D scene representation methodology for brain CT imaging. We have adapted the baseline Gaussian Splatting model to use suitable sparsity regularizers and adjusted the point initialization for brain CT radiographs, also removing the necessity of SfM methodologies in the process. We have tested our methodology leveraging 20 sets of DRR generated using PPMI brain images as phantoms for the simulations. Our results demonstrate that our model outperforms the current state-of-the-art methodologies in the field. \\
Our work leaves many openings for further research. An essential next step involves writing a new type of camera closely matching the detector array in CT imaging which is based on a curved orthogonal plane. This approximation would be much more accurate than our current pinhole camera approximation. Also, it is worth investigating and adapting the edge detection and feature extraction in SfM methodologies to accurately define the initial point cloud. Finally, we are planning to investigate the potential of using a latent representation of multiple medical scans using Gaussian Splatting representations.

\section{Acknowledgements}

We have used brain CT open-access datasets made available by PPMI. To synthesize DRR for our experiments, we used software provided by Plastimatch, an open source software for medical image computation. Finally, we based our model implementation on Gaussian Splatting by INRIA. 

\pagebreak

%
%
%
%

\bibliography{references}
\bibliographystyle{splncs04}
%

\pagebreak

\end{document}